\documentstyle[amssymb,preprint,aps]{revtex}

\begin{document}
\title{Hybrid Inflation in Supergravity with
$\left( SU(1,1)/U(1)\right) ^{m}$ K$%
\ddot{a}$hler Manifolds}
\author{C. Panagiotakopoulos}
\address{Theory Division, CERN, CH-1211 Geneva 23, Switzerland\\
and\\
Physics Division, School of Technology, Aristotle University of\\
Thessaloniki, 54006 Thessaloniki, Greece}
\maketitle

\begin{abstract}
In the presence of fields without superpotential but with large vevs through
D-terms the mass-squared of the inflaton in the context of supergravity
hybrid inflation receives positive contributions which could cancel the
possibly negative K$\ddot{a}$hler potential ones. The mechanism is
demonstrated using K$\ddot{a}$hler potentials associated with products of $%
SU(1,1)/U(1)$ K$\ddot{a}$hler manifolds. In a particularly simple model of
this type all supergravity corrections to the F-term potential turn out to
be proportional to the inflaton mass allowing even for an essentially
completely flat inflationary potential. The model also allows for a
detectable gravitational wave contribution to the microwave background
anisotropy. Its initial conditions are quite natural largely due to a built
in mechanism for a first stage of ``chaotic'' D-term inflation.
\end{abstract}

\newpage

The hybrid inflationary scenario \cite{linde94} has many advantages compared
to most other inflationary models \cite{linde90}. It does not involve tiny
inflaton self-couplings and succeeds in reconnecting inflation with phase
transitions in grand unified theories (GUTs). In its simplest realization it
involves a gauge singlet inflaton and a possibly gauge non-singlet
non-inflaton field. During inflation the non-inflaton field finds itself
trapped in a false vacuum state and the universe expands quasi-exponentially
dominated by the almost constant false vacuum energy density. Inflation ends
with (or just before) a very rapid phase transition when the non-inflaton
field rolls to its true vacuum state (``waterfall'').

The simplest supersymmetric (SUSY) particle physics model implementing the
above scenario in the context of a ``unifying'' gauge group $G$ 
(of rank $\geq 5)$ which
breaks spontaneously directly to the Standard Model (SM) gauge group $G_{S}$ 
$\equiv SU(3)_{c}\times SU(2)_{L}\times U(1)_{Y}$ at a scale $M_{X}\sim
10^{16}\>GeV$ is described by a superpotential which includes the terms \cite
{dvali} 
\begin{equation}
W=S(-\mu ^{2}+\lambda \Phi \bar{\Phi}).
\end{equation}
Here $\Phi$, $\bar{\Phi}$ is a conjugate pair of left-handed SM singlet
superfields which belong to $N_{d}$-dimensional representations of $G$ and
reduce its rank by their vacuum expectation values (vevs), $S$ is a gauge
singlet left-handed superfield, $\mu $ is a superheavy mass scale related to 
$M_{X}$ and $\lambda $ a real and positive coupling constant. The
superpotential terms in Eq. (1) are the dominant couplings involving the
superfields $S$, $\Phi$, $\bar{\Phi}$ which are consistent with a
continuous R-symmetry under which $W\to e^{i\gamma }W$, $S\to e^{i\gamma }S$%
, $\Phi $ $\to \Phi $ and $\bar{\Phi}\to \bar{\Phi}$. The potential obtained
from $W$ has, in the supersymmetric limit, a SUSY vacuum at 
\begin{equation}
<S>=0,<\Phi ><\bar{\Phi}>=\frac{\mu ^{2}}{\lambda }=\left( \frac{M_{X}}{g}%
\right) ^{2},\>\>\mid <\Phi >\mid =\mid <\bar{\Phi}>\mid ,
\end{equation}
where the scalar components of the superfields are denoted by the same
symbols as the corresponding superfields, $M_{X}$ is the mass acquired by
the gauge bosons and $g$ the $G$ gauge coupling constant. By an appropriate
R-trasformation we can bring the complex field $S$ on the real axis, i.e. $%
S\equiv \frac{1}{\sqrt{2}}\sigma $, where $\sigma $ is a real scalar field.
For any fixed value of $\sigma ^{2}>\sigma _{c}^{2}$, where $\sigma _{c}=%
\sqrt{2}\mu /\sqrt{\lambda }$, the potential of the globally supersymmetric
model has a minimum at $\Phi $ $=\bar{\Phi}=0$ with a $\sigma $-independent
value $V_{gl}=\mu ^{4}$ and the universe expands quasi-exponentially. When $%
\sigma ^{2}$ falls below $\sigma _{c}^{2}$ the false vacuum state at $\Phi%
=\bar{\Phi}=0$ becomes unstable and $\Phi$, $\bar{\Phi}$ roll rapidly to
their true vacuum.

If global supersymmetry is promoted to local one expects that the potential
will become very steep and an effective mass for the inflaton, large
compared to the inflationary Hubble parameter $H,$ will be generated
forbidding inflation even at small inflaton field values. In our case we can
investigate the consequences that $N=1$ supergravity has on our simple
hybrid inflationary model by restricting ourselves to the inflationary
trajectory ($\Phi $ $=\bar{\Phi}=0$). Then, we are allowed to use the simple
superpotential 
\begin{equation}
W=-\mu ^{2}S
\end{equation}
involving just the gauge singlet superfield $S.$ [Throughout our subsequent
discussion we make use of units in which the reduced Planck scale $%
m_{Pl}\equiv M_{Pl}/\sqrt{8\pi }\simeq 2.4355\times 10^{18}\>GeV\>\>$ is
equal to $1$ $(M_{Pl}\simeq 1.221\times 10^{19}\>GeV$ is the Planck mass).]
If the minimal K$\ddot{a}$hler potential $K=\left| S\right| ^{2}\ $leading
to canonical kinetic terms for $\sigma $ is employed the ``canonical''
potential $V_{can}$ acquires a slope and becomes \cite{cope} \cite{pan} \cite
{linde97} 
\begin{equation}
V_{can}=\mu ^{4}\left( 1-\left| S\right| ^{2}\ +\left| S\right| ^{4}\right) {%
e^{\left| S\right| ^{2}\ }=\mu }^{4}\sum_{k=0}^{\infty }\frac{(k-1)^{2}}{k!}%
\left| S\right| ^{2k}.
\end{equation}
$V_{can}$ of Eq. (4) does not allow inflation unless $\left| S\right| ^{2}\
\ll 1.$ From its expansion as a power series in $\left| S\right| ^{2}\ $we
see that, due to an ``accidental'' cancellation, the linear term in $\left|
S\right| ^{2}\ $is missing and no mass-squared term is generated for $\sigma 
$. Small deviations from the minimal form of the K$\ddot{a}$hler potential
respecting the R-symmetry lead to a K$\ddot{a}$hler potential \cite{pan1} 
\begin{equation}
K=\left| S\right| ^{2}\ -\frac{\alpha }{4}\left| S\right| ^{4}+\ldots .\text{
}
\end{equation}
which, in turn, gives rise to a potential admitting an expansion 
\begin{equation}
V=\mu ^{4}\left( 1+\alpha \left| S\right| ^{2}\ +\ldots \right) \qquad
\left( \left| S\right| ^{2}\ <<1\right)
\end{equation}
in which a linear term in $\left| S\right| ^{2}\ $proportional to the small
parameter $\alpha $ is now generated. All higher powers of $\left| S\right|
^{2}\ $are still present in the series with coefficients only slightly
different from the corresponding ones of Eq. (4). Almost all studies of
hybrid inflation in the context of the simplest model of Eq. (1) rely,
explicitly or implicitly, on this ``miraculous'' cancellation of the mass of 
$\sigma $ in $N=1$ canonical supergravity.

Our discussion so far seems to indicate that the only potential source of
mass for $\sigma $ is the next to leading term in the expansion of the K$%
\ddot{a}$hler potential in powers of $\left| S\right| ^{2}\ $which must have
a small and negative coefficient. This conclusion is certainly correct if
all other fields are assumed to play absolutely no role during inflation. In
the present paper we show that fields which do not contribute to the
superpotential and are singlets under the ``unifying'' group $G$, and
therefore could be regarded as ``playing no role'', do contribute to the
mass-squared of $\sigma $ if they acquire large vevs. Then, in the presence
of such fields the ``miraculous'' properties of the minimal K$\ddot{a}$hler
potential are lost \footnote{%
$ $The danger that the ``miraculous'' cancellation of the inflaton mass might
be destroyed by fields acquiring large vevs has also been pointed out in 
\cite{lr}.} $ $but other ``miraculous'' cancellations might occur involving
other, possibly better motivated, types of K$\ddot{a}$hler potentials.

Let us consider a $G$-singlet chiral superfield $Z$ which does not
contribute to the superpotential at all because, for instance, it has
non-zero charge, let us say $-1,$ under an ``anomalous'' $U(1)$ gauge
symmetry and, as we assume, all other superfields which have a $U(1)$ charge
can be safely ignored. Also let us assume that $\frac{\partial ^{2}K}{\partial%
S\partial Z^{*}}=\frac{\partial ^{2}K}{\partial Z\partial S^{*}}=0$ such
that $K=K_{1}(\left| S\right| ^{2})+K_{2}(\left| Z\right| ^{2}).$
Then, with the
parameters $\mu $ and $\lambda $ in $W$ renamed as $\mu ^{^{\prime }}$ and $%
\lambda ^{^{\prime }}$, the scalar potential becomes 
\begin{equation}
V=\mu ^{^{\prime }4}\left\{ \left| 1+S\frac{\partial K}{\partial S}\right|
^{2}\left( \frac{\partial ^{2}K}{\partial S\partial S^{*}}\right)
^{-1}+\left( \left| \frac{\partial K}{\partial Z}\right| ^{2}\left( \frac{%
\partial ^{2}K}{\partial Z\partial Z^{*}}\right) ^{-1}-3\right) \left|
S\right| ^{2}\ \right\} e^{K}+\frac{1}{2}g_{1}^{2}\left( \frac{\partial K}{%
\partial Z}Z-\xi \right) ^{2},
\end{equation}
where the first(second) term is the F(D)-term,
 $\xi >0$ is a Fayet-Iliopoulos term and $g_{1}$ the gauge coupling of
the ``anomalous'' $U(1)$ gauge symmetry. Minimization of such a potential
for fixed $\left| S\right| ^{2}$ not much larger than unity, assuming $%
\left| S\right| ^{2}\ $takes values away from any points where the potential
is singular and $\mu ^{^{\prime }2}\ll \xi $, typically gives rise to a $<$ $%
\mid Z\mid ^{2}>\equiv v^{2}\sim \xi $ with $\left( \left| \frac{\partial K}{%
\partial Z}\right| ^{2}\left( \frac{\partial ^{2}K}{\partial Z\partial Z^{*}}%
\right) ^{-1}\right) _{\left| Z\right| =v}\sim v^{2}\sim \xi $ and therefore
a contribution to the mass-squared of $\sigma $ 
\begin{equation}
\delta m_{\sigma}^{2}=\left( \left| \frac{\partial K}%
{\partial Z}\right| ^{2}\left( 
\frac{\partial ^{2}K}{\partial Z\partial Z^{*}}\right) ^{-1}\right) _{\left|
Z\right| =v}\mu ^{^{\prime }4}e^{K_{2}(v^{2})}
\end{equation}
of the order of $\xi $ in units of the false vacuum energy density. For the
sake of convenience we absorb the factor $e^{K_{2}(v^{2})}$ appearing in the
F-term potential in the reintroduced parameters $\mu =\mu ^{^{\prime
}}e^{K_{2}(v^{2})/4}$ and $\lambda =\lambda ^{^{\prime }}e^{K_{2}(v^{2})/2}$
obeying the relation $\frac{\mu }{\sqrt{\lambda }}=\frac{\mu ^{^{\prime }}}{%
\sqrt{\lambda ^{^{\prime }}}}$.

It would be very interesting if the contribution of $Z$ to the mass-squared
of $\sigma $ in units of the false vacuum energy density were independent of
the value of $Z$. This is exactly the case if $Z$ enters the K$\ddot{a}$hler
potential through a function $K_{2\text{ }}$ of the ``no-scale'' type \cite
{crem} \cite{el} 
\begin{equation}
K_{2}=-n\ln \left| f(Z)+f^{*}(Z^{*})\right| ,
\end{equation}
where $n$ is an integer, $f(Z)$ an analytic function of $Z$ and $%
f^{*}(Z^{*}) $ its complex conjugate which is a function of $Z^{*}$. The
corresponding K$\ddot{a}$hler manifold is an Einstein-K$\ddot{a}$hler
manifold of constant scalar curvature $2/n$ with a non-compact $SU(1,1)$
global symmetry and a local $U(1)$ symmetry, namely the coset space $%
SU(1,1)/U(1)$. Different choices of $f(Z)$ correspond to $Z$ field
redefinitions. Since, however, we want $Z$ to transform non-trivially under
a continuous $U(1)$ symmetry the function $f(Z)$ must be a linear function
of ln$Z$. Thus, without loss of generality, we are led to the K$\ddot{a}$%
hler potentials 
\begin{equation}
K_{2}(\left| Z\right| ^{2})=-n\ln \left( -\ln \left| Z\right| ^{2}\right)
\qquad \left(0< \left| Z\right| ^{2}<1\right) .
\end{equation}
Such a choice makes the contribution of $Z$ to the mass-squared of $\sigma $%
\begin{equation}
\delta m_{\sigma}^{2}=n\mu ^{4},
\end{equation}
an integer multiple of the false vacuum energy density.

In order for the above discussion to be of any use we should, of course,
find K$\ddot{a}$hler potentials $K_{1}(\left| S\right| ^{2})$ whose
expansion in powers of $\left| S\right| ^{2}$ has a positive next to leading
term. A class of such K$\ddot{a}$hler potentials is given by 
\begin{equation}
K_{1}(\left| S\right| ^{2})=-N\ln \left( 1-\frac{\left| S\right| ^{2}}{N}%
\right) \qquad \left( \left| S\right| ^{2}<N\right) ,
\end{equation}
where $N$ is an integer. The corresponding K$\ddot{a}$hler manifold is again
the coset space $SU(1,1)/U(1)$ with constant scalar curvature $2/N.$
[Actually the kinetic terms of $Z$ and $S$ with K$\ddot{a}$hler potentials
given by Eqs. (9) and (12), respectively (with $n=N$) transform into each
other under the redefinition $f(Z)=\frac{\sqrt{N}+S}{\sqrt{N}-S}.$]
Expanding $K_{1}$ of Eq. (12) in powers of $\left| S\right| ^{2}$%
\begin{equation}
K_{1}(\left| S\right| ^{2})=\left| S\right| ^{2}+\frac{1}{2N}\left| S\right|
^{4}+\ldots
\end{equation}
and comparing with Eqs. (5) and (6) we see that the mass-squared of $\sigma $
is 
\begin{equation}
m_{\sigma}^{2}=\left( -\frac{2}{N}+\left| \frac{\partial K}{\partial Z}%
\right|^{2}\left( \frac{\partial ^{2}K}{\partial Z\partial Z^{*}}\right)
^{-1}\right) _{\left| Z\right| =v}\mu ^{4}\equiv \beta \mu ^{4}.
\end{equation}
For all $N$ we can make $m_{\sigma}^{2}$ positive (or, by fine tuning, zero)
 through
appropriately chosen vevs ($\xi $ parameters) of $Z$-type fields. The most
interesting cases, however, occur for $N=1$ or $N=2$ because $2/N$ is an
integer and the option of naturally making $m_{\sigma}^{2}$
 vanish by employing $Z$%
-type fields with K$\ddot{a}$hler potentials given by Eq. (10) (with $n=2$
or $n=1,$ respectively) becomes now available. A small positive
 $m_{\sigma}^{2}$
could be subsequently generated through additional $Z$-type fields which
acquire vevs of the order of appropriately chosen $\xi $ parameters.

It is important to realize that $\beta \equiv m_{\sigma}^{2}/\mu ^{4}$ is
approximately constant only for $\left( 1-\frac{\left| S\right| ^{2}}{N}%
\right) ^{N}\gg \mu ^{4}/g_{1}^{2}\xi .$ As $\left( 1-\frac{\left| S\right|
^{2}}{N}\right) ^{N}$ approaches zero the $Z$ fields get displaced from
their vevs at small $\left| S\right| ^{2}$ and $\beta $ will eventually
vanish. Further decrease of $\left( 1-\frac{\left| S\right| ^{2}}{N}\right)
^{N}$ will result in a negative $\beta .$ This can be avoided in the cases $%
N=1,2$ provided the cancellation mechanism involving the K$\ddot{a}$hler
potentials of Eq. (10) is used.

The choices $N=1$ or $N=2$ deserve further study because, as we now show, in
these cases all supergravity corrections to the F-term potential are
proportional to the mass-squared $m_{\sigma}^{2}$ of the field $\sigma $ or,
equivalently, to the parameter $\beta .$ This offers the possibility of
suppressing or even eliminating all supergravity corrections to the
inflationary trajectory by suppressing or making vanish the parameter $\beta 
$. Indeed, a straightforward substitution of the K$\ddot{a}$hler potential $%
K_{1}(\left| S\right| ^{2})$ of Eq. (12) in Eq. (7) gives 
\begin{equation}
V\simeq \mu ^{4}\left\{ \left( 1-\left| S\right| ^{2}\ +\frac{(N-1)^{2}}{%
N^{2}}\left| S\right| ^{4}\right) +\beta \left| S\right| ^{2}\ \right\}
\left( 1-\frac{\left| S\right| ^{2}}{N}\right) ^{-N}
\end{equation}
(up to terms $\sim \mu ^{8}$) which, for $N=1,2$ only, becomes 
\begin{equation}
V\simeq \mu ^{4}\left\{ 1+\beta \left| S\right| ^{2}\ \left( 1-\frac{\left|
S\right| ^{2}}{N}\right) ^{-N}\right\} \qquad \left( N=1,2\right)
\end{equation}
independently of the mechanism chosen to make $\beta \geq 0.$ In
particular, the combinations of $K_{1}(\left| S\right| ^{2})$ of Eq. (12)
with $N=1$ and $K_{2}(\left| Z\right| ^{2})$ of Eq. (10) with $n=2$ (or two $%
Z$-type fields each entering a $K_{2}$ with $n=1$) or $K_{1}(\left| S\right|
^{2})$ with $N=2$ and $K_{2}(\left| Z\right| ^{2})$ with $n=1$ give $\beta
=0 $ and consequently a completely flat potential.

Because of the possibility of suppressing the supergravity corrections
through a single parameter $\beta $ we expect that the models with $N=1,2$
will lead to some novel features compared to the quasi-canonical case \cite
{pan1}, mostly due to the anticipated extension of inflation to $\sigma $
values close or even slightly larger than unity. In particular, one might
hope for the possibility of predicting a detectable gravitational wave
signal in the cosmic microwave background anisotropy which requires an
inflaton field variation of order unity \cite{ly}. In order to decide which
of the two models ($N=1,2$) has a flatter potential and consequently a
better chance for novel features we express their potentials in terms of the
canonically normalized inflaton field $\sigma _{infl}$. The relation between 
$\sigma $ and $\sigma _{infl}$ is 
\begin{equation}
\frac{\sigma }{\sqrt{2N}}=\tanh \frac{\sigma _{infl}}{\sqrt{2N}}\text{ }.
\end{equation}
Then, the potential of the $N=1$ model becomes 
\begin{equation}
V\simeq \mu ^{4}\left[ 1+\frac{\beta }{2}\left( \cosh \left( \sqrt{2}\sigma
_{infl}\right) -1\right) \right] =\mu ^{4}\left[ 1+\frac{\beta }{2}%
\sum_{k=1}^{\infty }\frac{2^{k}}{(2k)!}\sigma _{infl}^{2k}\right] \qquad
\left( N=1\right) ,
\end{equation}
whereas the potential of the $N=2$ model becomes 
\begin{equation}
V\simeq \mu ^{4}\left[ 1+\frac{\beta }{4}\left( \cosh \left( 2\sigma
_{infl}\right) -1\right) \right] =\mu ^{4}\left[ 1+\frac{\beta }{2}%
\sum_{k=1}^{\infty }\frac{2^{2k-1}}{(2k)!}\sigma _{infl}^{2k}\right] \qquad
\left( N=2\right) .
\end{equation}
Comparing the coefficients of the corresponding terms of the series in Eqs.
(18) and (19) we clearly conclude that the model with $N=1$ has flatter
potential and is more promising.

In the following we concentrate on the model with $N=1.$ The derivative of
the potential with respect to $\sigma _{infl}$ expressed as a function of $%
\sigma $ is 
\begin{equation}
V^{\prime }=\frac{\mu ^{4}}{2\sigma }\left[ N_{d}\left( \frac{\lambda }{2\pi 
}\right) ^{2}\left( 1-\frac{\sigma ^{2}}{2}\right) +2\beta \sigma ^{2}\left(
1-\frac{\sigma ^{2}}{2}\right) ^{-1}\right] ,
\end{equation}
where the effect of the radiative corrections \cite{dvali} due to the
conjugate pair of non-inflaton fields $\Phi$, $\bar{\Phi}$ belonging to $%
N_{d}$-dimensional representations of the gauge group $G$ are included.
Their vev is fixed at the Minimal Supersymmetric Standard Model (MSSM) value 
$\mu /\sqrt{\lambda }=M_{X}/g\simeq 0.011731$ ($M_{X}\simeq 2\times 10^{16}$ 
$GeV,g\simeq 0.7$) and correspondingly the critical value of $\sigma $ is $%
\sigma _{c}\simeq 0.01659.$

For the quadrupole anisotropy $\frac{\Delta T}{T}$ we employ the standard
formula \cite{lid}

\begin{equation}
\left( \frac{\Delta T}{T}\right) ^{2}\simeq \frac{1}{720\pi ^{2}}\left[ 
\frac{V^{3}}{V^{\prime 2}}+6.9V\right] _{\sigma _{H}},
\end{equation}
where $\sigma _{H}$ is the value of $\sigma $ when the scale $\ell _{H}$,
corresponding to the present horizon, crossed outside the inflationary
horizon. The first term in Eq. (21) is the scalar component $\left( \frac{%
\Delta T}{T}\right) _{S}^{2}$ of $\left( \frac{\Delta T}{T}\right) ^{2}$
whereas the second is the tensor one $\left( \frac{\Delta T}{T}\right)
_{T}^{2}$ which represents the gravitational wave contribution. Their ratio $%
r$ is 
\begin{equation}
r\equiv \left( \frac{\Delta T}{T}\right) _{T}^{2}/\left( \frac{\Delta T}{T}%
\right) _{S}^{2}\simeq 6.9\left( \frac{V^{\prime }}{V}\right) _{\sigma
_{H}}^{2}\simeq 27.6\left( \frac{\beta \sigma _{H}}{2-(1-\beta )\sigma
_{H}^{2}}\right) ^{2},
\end{equation}
assuming, as it turns out to be the case, that the effect of radiative
corrections is negligible at the point where the spectrum of temperature
fluctuations is normalized.

The number of e-foldings $\Delta N(\sigma _{in},\sigma _{f})$ for the time
period that $\sigma $ varies between the values $\sigma _{in}$ and $\sigma
_{f}\>\ (\sigma _{in}>\sigma _{f})$ is given, in the slow roll
approximation, by 
\begin{equation}
\Delta N(\sigma _{in},\sigma _{f})=-\int_{\sigma _{in}}^{\sigma {_{f}}}\frac{%
V}{V^{\prime }}\left( 1-\frac{\sigma ^{2}}{2}\right) ^{-1}d\sigma \text{.}
\end{equation}
The evaluation of the above integral is straightforward but the result is
rather lengthy and will not be given.

Let $\ell _{H}$ be the scale corresponding to our present horizon and $\ell
_{o}$ another length scale. Also let $\sigma _{o}$ be the value that $\sigma 
$ had when $\ell _{o}$ crossed outside the inflationary horizon. We define
the average spectral index $n(\ell _{o})$ for scales from $\ell _{o}$ to $%
\ell _{H}$ as 
\begin{equation}
n(\ell _{o})\equiv 1+2ln\left[ \left( \frac{\delta \rho }{\rho }\right)
_{\ell _{o}}/\left( \frac{\delta \rho }{\rho }\right) _{\ell _{H}}\right]
/ln\left( \frac{\ell _{H}}{\ell _{o}}\right) =1+2ln\left[ \left( \frac{%
V^{3/2}}{V^{\prime }}\right) _{\sigma _{o}}/\left( \frac{V^{3/2}}{V^{\prime }%
}\right) _{\sigma _{H}}\right] /\Delta N(\sigma _{H},\sigma _{o}).
\end{equation}
Here $(\delta \rho /\rho )_{\ell }$ is the amplitude of the energy density
fluctuations on the length scale $\ell $ as this scale crosses inside the
postinflationary horizon and $\Delta N(\sigma _{H},\sigma _{o})=ln(\ell
_{H}/\ell _{o})$.

It should be clear that all quantities characterizing inflation depend on
just two parameters, namely $\mu $ and $\beta .$ Therefore for each value of 
$\mu $ we can determine $\sigma _{H}\ (>0)$ and $\beta $ by normalizing the
spectrum according to Eq. (21) and requiring that $N_{H}$ $\equiv \Delta
N(\sigma _{H},\sigma _{c})$ takes the appropriate value. We choose $\frac{%
\Delta T}{T}=6.6\times 10^{-6}$ and $N_{H}\simeq 50$.

Table 1 gives the values of $\beta ,$ $\sigma _{H},$ $\sigma _{infl_{H}},$ $%
n\equiv n(\ell _{1})$, $n_{COBE}\equiv n(\ell _{2})$ and $r,$ where $\ell
_{1}$ $(\ell _{2})$ is the scale that corresponds to $1$ $Mpc$ $(2000$ $Mpc)$
today, for different values of $\mu $ assuming that the present horizon size
is $12000$ $Mpc$ and $N_{d}=1$. Comparing with the quasi-canonical case \cite
{pan1} we see clearly an increase in the values of $\beta .$ The spectrum of
density perturbations is again blue but the spectral index, which here also
shows a strong scale dependence, is severely lowered. As a result of this
drastic decrease of the spectral index the parameter $\mu $ now could be,
for the first time, very close to the MSSM gauge coupling unification scale $%
M_{X}.$ This has as a consequence a dramatic increase of $r$ which reaches
values that could be detectable in the foreseeable future. As already
mentioned all these effects are due to the suppression of the supergravity
corrections which allows inflation to take place at larger values of the
inflaton field and makes our model approach the original hybrid model \cite
{linde94}.

Table 2 gives, for fixed $\mu =7.3\times 10^{-3}$, the values of $\beta ,$ 
$\sigma _{H},$ $\sigma _{infl_{H}},$ $n$, $n_{COBE}$ and $r$ as functions of
the dimensionality $N_{d}$ of the representations to which $\Phi$, $\bar{%
\Phi}$ belong. We see that as $N_{d}$ increases $\beta $ and the spectral
index decrease whereas $\sigma _{H}$ and $\sigma _{infl_{H}}$ increase.
Remarkably enough $r$ remains essentially unaltered.

As $\sigma _{infl}$ $(>0)$ increases the constant term in $V$ of Eq. (18)
becomes gradually irrelevant and the potential along the inflationary
trajectory can be approximated, for $\sigma _{infl}^{c}\gtrsim \sigma
_{infl}\gg 1$ (with $\sigma _{infl}^{c}$ a model-dependent critical value),
by the well-known one \footnote{%
$ $The possibility that the evolution of the universe before the onset of
``slow-roll'' inflation could be described by potentials of this type has
been considered in \cite{lr}.} 
\begin{equation}
V=V_{0}e^{\sqrt{2}\sigma _{infl}}\text{ }\qquad (\sigma _{infl}^{c}\gtrsim
\sigma _{infl}\gg 1)
\end{equation}
(with $V_{0}=\mu ^{4}\beta /4)$. The equation of motion of $\sigma
_{infl}(t) $ with such a potential in
a $\sigma _{infl}$-dominated universe admits a solution with 
\begin{equation}
\frac{d}{dt}\sigma _{infl}=-\sqrt{V}=-\mu ^{2}\frac{\sqrt{\beta }}{2}e^{%
\frac{\sigma _{infl}}{\sqrt{2}}}\text{ }\qquad (\sigma _{infl}^{c}\gtrsim
\sigma _{infl}\gg 1),
\end{equation}
meaning that eventually $\sigma _{infl}$ acquires a ``terminal'' velocity
which is a function of $\sigma _{infl}$ independently of the initial
conditions. Then, the scale factor $R(t)$ of the universe expands like $%
R\sim t$ and the energy density $\rho $ falls like $\rho \sim R^{-2}$ or,
equivalently, the Hubble length $H^{-1}\sim R.$ Thus, although the expansion
is certainly slower than inflation it leaves invariant the volume of any
space region in units of $H^{-3}$ thereby leading to the onset of inflation
at $\rho \sim \mu ^{4}\ll 1.$ This will happen provided the inflationary
trajectory extends almost up to $\rho \sim 1$ and a natural set of initial
conditions develops sufficiently early into a set of field values and
velocities describing motion on the inflationary trajectory.

To address these issues we need to further specify the model. We choose to
introduce, in addition to the superfields $S$, $\Phi$, $\bar{\Phi}$, two $G$%
-singlet superfields $Z_{1},Z_{2}$ with charges $(-1,0)$ and $(0,-q)$,
respectively (with $q>0$) under two ``anomalous'' $U(1)$ gauge symmetries.
The K$\ddot{a}$hler potential is chosen to be 
\begin{equation}
K=-\ln \left( 1-\left| S\right| ^{2}\right) -2\ln \left( -\ln \left|
Z_{1}\right| ^{2}\right) -\ln \left( 1-\left| Z_{2}\right| ^{2}\right)
+\left| \Phi \right| ^{2}+\left| \bar{\Phi}\right| ^{2}
\end{equation}
($ \left| S\right| ^{2},\left| Z_{2}\right| ^{2} <1$,
$ 0< \left| Z_{1}\right|^{2}<1$) with the superpotential being 
$W=S(-\mu ^{^{\prime}2}+\lambda ^{^{\prime }}\Phi \bar{\Phi}).$

To discuss the (extension of the) inflationary trajectory at large $\left|
\sigma _{infl}\right| $ values we ignore the superfields $\Phi$, $\bar{\Phi}
$ and we obtain the potential 
\[
\text{ }V=\mu ^{^{\prime }4}\frac{\xi _{1}^{2}}{4}e^{2\zeta }\left[ 1+\frac{1%
}{4}\left( \cosh \left( \sqrt{2}\chi \right) -1\right) \left( \cosh \left( 
\sqrt{2}\sigma _{infl}\right) +1\right) \right] 
\]
\begin{equation}
+\frac{1}{2}g_{1}^{2}\xi _{1}^{2}\left( e^{\zeta }-1\right) ^{2}+\frac{1}{2}%
g_{2}^{2}\left[ \frac{q}{2}\left( \cosh \left( \sqrt{2}\chi \right)
-1\right) -\xi _{2}\right] ^{2},
\end{equation}
where $g_{1},$ $g_{2}\sim 1$ are the gauge couplings of the ``anomalous'' $%
U(1)$' s and $\xi _{1}\sim 1,$ $\xi _{2}\sim 10^{-1}-10^{-2}$ the
corresponding (positive) Fayet-Iliopoulos terms. The canonically normalized
real scalar fields $\zeta $ and $\chi $ are defined through the relations 
\begin{equation}
e^{-\zeta }\equiv -\frac{\xi _{1}}{2}\ln \left| Z_{1}\right| ^{2},\text{ }%
\tanh \frac{\chi }{\sqrt{2}}\equiv 
\mathop{\rm Re}%
Z_{2},
\end{equation}
with the complex scalar fields $Z_{1},Z_{2}$ brought to the real axis by
gauge transformations. For fixed $\left| \sigma _{infl}\right| <\sigma
_{infl}^{c},$ where 
\begin{equation}
\sigma _{infl}^{c}=\frac{1}{\sqrt{2}}\arccos h\left[ \frac{8g_{2}^{2}\xi
_{2}q}{\mu ^{^{\prime }4}\xi _{1}^{2}}\left( 1+\frac{\mu ^{^{\prime }4}}{%
2g_{1}^{2}}\right) ^{2}-1\right] ,
\end{equation}
$V$ is minimized at $\zeta =\zeta _{\min }<0$ with $\left| \zeta _{\min
}\right| \ll 1$ and $\chi =\chi _{\min }$ with 
\begin{equation}
\chi _{\min }\simeq \frac{1}{\sqrt{2}}\arccos h\left[ 1+\frac{2\xi _{2}}{q}%
\left( 1-\frac{\cosh \left( \sqrt{2}\sigma _{infl}\right) +1}{\cosh \left( 
\sqrt{2}\sigma _{infl}^{c}\right) +1}\right) \right] .
\end{equation}
If $\left| \sigma _{infl}\right| \ll \sigma _{infl}^{c}$ its value at the
minimum
is given by Eq. (18) with $\beta \simeq \frac{\xi _{2}}{q}\left( 1+\frac{\xi
_{2}}{q}\right) ^{-1}$ and $\mu ^{4}\simeq \mu ^{^{\prime }4}\frac{\xi
_{1}^{2}}{4}\left( 1+\frac{\xi _{2}}{q}\right) .$[Also the coupling $\lambda 
$ is related to the coupling $\lambda ^{^{\prime }}$ appearing in $W$
through the relation $\lambda \simeq \lambda ^{^{\prime }}\frac{\xi _{1}}{2}%
\left( 1+\frac{\xi _{2}}{q}\right) ^{\frac{1}{2}}$ such that $\frac{\mu }{%
\sqrt{\lambda }}=\frac{\mu ^{^{\prime }}}{\sqrt{\lambda ^{^{\prime }}}}=%
\frac{M_{X}}{g}.$] For $\left| \sigma _{infl}\right| \geqslant \sigma
_{infl}^{c}$ instead, $\zeta _{\min }=-\ln \left( 1+\frac{\mu ^{^{\prime }4}}{%
2g_{1}^{2}}\right) ,$ $\chi _{\min }=0$ and the potential along the
inflationary trajectory is completely flat 
\begin{equation}
V=\mu ^{^{\prime }4}\frac{\xi _{1}^{2}}{4}\left( 1+\frac{\mu ^{^{\prime }4}}{%
2g_{1}^{2}}\right) ^{-1}+\frac{1}{2}g_{2}^{2}\xi _{2}^{2}\simeq \frac{1}{2}%
g_{2}^{2}\xi _{2}^{2}\quad \left( \left| \sigma _{infl}\right| \geqslant
\sigma _{infl}^{c}\right) 
\end{equation}
(neglecting the tiny radiative corrections). Thus, the inflationary
trajectory could provide a description of the evolution of the universe only
after $\rho \lesssim \frac{1}{2}g_{2}^{2}\xi _{2}^{2}\sim 10^{-2}-10^{-4}$
provided, of course, the initial conditions are appropriate. If this is the
case the universe undergoes a huge period of D-term inflation \cite{dt},
essentially governed by quantum fluctuations, along the flat region of the
potential followed by the expansion described approximately by the potential
of Eq. (25) before the ``observable'' inflation takes place.

An investigation of the initial conditions necessarily involves the fields $%
\Phi$, $\bar{\Phi}.$ To simplify the discussion we keep only one out of the
four real scalars in $\Phi$, $\bar{\Phi}$ by setting $\Phi $ $=\bar{\Phi}=%
\frac{\varphi }{2}$, where $\varphi $ is a canonically normalized real scalar
field, and we consider a trancated version of the complete
scalar potential 
\begin{equation}
V=\frac{\lambda ^{2}}{4}\varphi ^{2}\left( \cosh \left( \sqrt{2}\sigma
_{infl}\right) -1\right) e^{2\zeta }+\frac{1}{2}g_{1}^{2}\xi _{1}^{2}\left(
e^{\zeta }-1\right) ^{2}
\end{equation}
(technically justified for $\frac{\mu ^{2}}{\lambda }\ll \varphi
^{2}\lesssim 10^{-1}$ and $\chi ^{2}\simeq 2\beta $) possessing all its
salient features. We assume that initially 
$\left| \sigma_{infl_{0}}\right|\gg 1$,
$e^{\zeta_{0}}\ll 1$, $\varphi_{0}^{2}\sim 10^{-1}$
and the initial time derivatives of all fields vanish.  
Notice that $e^{\zeta _{0}}\ll 1$ is required
in order for $\rho _{0}\lesssim 1$ if $\left| \sigma _{infl_{0}}\right|$
is sufficiently large. Then, $e^{\zeta}$ starts decreasing further unless
the F-term in Eq. (33) is smaller than $\rho_{0}e^{\zeta_{0}}$ to begin with.
To ensure a sufficiently fast decrease of $\varphi ^{2}$ we assume that
$\frac{\partial ^{2}V}{\partial \varphi ^{2}}\gtrsim \rho $ holds from
the beginning which, for the initial conditions adopted, translates into 
\begin{equation}
\left( \cosh \left( \sqrt{2}\sigma _{infl_{0}}\right) -1\right) e^{2\zeta
_{0}}\gtrsim \frac{g_{1}^{2}\xi _{1}^{2}}{\lambda ^{2}}  .
\end{equation}
With $\varphi ^{2}$ decreasing fast the relation  
$\frac{1}{V}\frac{\partial V}{\partial \zeta }\simeq \frac{1}{V}%
\frac{\partial ^{2}V}{\partial \zeta ^{2}}\simeq -2e^{\zeta \text{ }}$ ($%
e^{\zeta }\ll 1$) is soon established
and the universe experiences a stage of ``chaotic''
D-term inflation with $H=H_{1}\simeq \frac{1}{\sqrt{6}}g_{1}\xi
_{1}(1-e^{\zeta })$ which begins when 
$\zeta=\zeta_{beg}\lesssim \zeta_{0}< 0.$ 
The total number of e-foldings $N_{tot}$ as $\zeta $
varies from $\zeta _{beg}$ towards its minimum at $\zeta _{\min }\simeq 0$ is 
\begin{equation}
N_{tot}\simeq \frac{1}{2}\left( e^{-\zeta _{beg}}-e\right)\gtrsim \frac{1}{2}
\left( e^{-\zeta_{0}}-e\right)
\end{equation}
(assuming inflation ends at $\zeta _{end}=-1$). Moreover, $\frac{\partial
^{2}V}{\partial \sigma _{infl}^{2}}\simeq \varphi ^{2}\frac{\partial ^{2}V}{%
\partial \varphi ^{2}}.$ Consequently, even if initially $\frac{\partial
^{2}V}{\partial \sigma _{infl}^{2}}\gtrsim \rho $ (i.e. $\left| \sigma
_{infl_{0}}\right| \gg 1$), very soon $\frac{\partial ^{2}V}{\partial \sigma
_{infl}^{2}}\ll \rho $ and $\left| \sigma _{infl}\right| $ stays large
with $\varphi ^{2}$ becoming very small. Thus, when the ``chaotic'' D-term
inflation is over the field configuration is close to the inflationary
trajectory but $\sigma _{infl}$ does not reach its terminal velocity as long
as $\rho $ is dominated by the field $\zeta $ which oscillates coherently
about its minimum. It is understood that $\chi $ (which plays a secondary
role in the beginning) will also oscillate about its minimum. Whether this
minimum is at $\chi =0$ or not depends on whether at the end of the
``chaotic'' D-term inflation $\left| \sigma _{infl}\right| $ is larger or
smaller than $\sigma _{infl}^{c}.$ In the former case a second stage of
D-term inflation with $H=H_{2}\simeq\frac{1}{\sqrt{6}}g_{2}\xi _{2}$ will take
place as discussed earlier. [Of course, $\left| \sigma _{infl}\right|
\geqslant \sigma _{infl}^{c}$ requires relatively large initial values of $%
\left| \sigma _{infl}\right| $ ($\left| \sigma _{infl_{0}}\right| >10$) and
large negative values of $\zeta _{0}.$] In the latter case, in which we 
concentrate, there will be no
second stage of D-term inflation. The Hubble length $H^{-1}$ will grow like $%
R^{\frac{3}{2}}$ as long as the universe is matter dominated and gradually
will switch to an expansion law $H^{-1}\sim R$, as discussed earlier.
Overall, between the ``chaotic'' D-term inflation and the ``observable''
inflation, $H^{-1}/R$ grows only during the era of matter domination like $%
R^{\frac{1}{2}}\sim \rho ^{-\frac{1}{6}}$ or by at most a factor $\sim 20-90$
for $\mu \sim (7.3-1)\times 10^{-3}.$ The effect of this growth on the onset
of ``observable'' inflation \cite{tet} can be ``compensated for'' if the
``homogeneous'' region of volume $\sim H_{1}^{-3}$ required for the onset
of the ``chaotic'' D-term inflation experiences at least $3-4.5$
e-foldings of expansion during this first Planck-scale inflationary stage 
\cite{pt} \cite{lt}.
Notice that $\rho _{0}\sim 1$ is
obtained without invoking enormous initial field values and therefore
suppression of the gradient energy density in the above ``homogeneous''
region is not unnatural.

So far we neglected the effects of the quantum fluctuations during the stage
of ``chaotic'' D-term inflation. These fluctuations generate at the end of
this first inflationary stage classical perturbations $\Delta \sigma
_{infl}\sim \frac{H_{1}}{2\pi }$ of the (nearly) massless field $\sigma
_{infl}$ and perturbations $\Delta \varphi $ $\sim c\frac{H_{1}^{2}}{%
m_{\varphi }}$ of the massive field $\varphi $ (with $H_{1}$ the Hubble
parameter towards the end of the ``chaotic'' D-term inflation, $c\sim
10^{-1} $ and $m_{\varphi }$ the mass of $\varphi $). $\Delta \sigma _{infl}$
gives rise to a spatial gradient energy density $\sim \frac{H_{1}^{4}}{4\pi
^{2}}$ which falls like $R^{-2}.$ Under the extreme assumption of matter
domination during the whole intermediate stage before the onset of
``observable'' inflation this gradient energy density remains subdominant
until $\rho \sim \mu ^{4}$ provided $\left( \frac{3H_{1}^{2}}{\mu }\right) ^{%
\frac{4}{3}}\ll 36\pi ^{2}.$ $\Delta \varphi $ gives rise to $\frac{\partial
^{2}V}{\partial \sigma _{infl}^{2}}\simeq \left( \Delta \varphi \right) ^{2}%
\frac{\partial ^{2}V}{\partial \varphi ^{2}}\sim c^{2}H_{1}^{4}$ at the end
of the first stage of inflation which, assuming constant $\sigma _{infl},$
falls like $R^{-3}$ during the subsequent expansion. Thus, the contribution
of the quantum fluctuations to $\frac{\partial ^{2}V}{\partial \sigma
_{infl}^{2}}$ is at most $\sim c^{2}H_{1}^{2}H^{2}\ll H^{2}$ and therefore
negligible.

A numerical investigation reveals that even more natural initial conditions
than the ones adopted in the above somewhat simplified discussion do lead to
evolution along the inflationary trajectory. Let us choose $q=1$ and $%
g_{1}=g_{2}=\xi _{1}=\frac{1}{\sqrt{2}}\simeq 0.7.$ Then, with $\Phi$, $%
\bar{\Phi}$ along the D-flat direction $\Phi $ $=\bar{\Phi}=\frac{1}{2}%
(\varphi $ $+i\psi ),$ where $\varphi $, $\psi $ are canonically
normalized real scalar fields, it is possible to set $\varphi _{0}=\psi
_{0}=\chi _{0}=1$ and $\zeta _{0}=-2.5$ (in order to obtain sufficient
expansion during the ``chaotic'' D-term inflation), assuming zero initial
time derivatives for these fields. Moreover, we start with $\frac{d}{dt}%
\sigma _{infl}=-1$ and determine $\sigma _{infl_{0}}$ ($>0$) by requiring
$\rho_{0}=1.$ For $\mu =7.3\times 10^{-3}$ and $\beta =0.033$ the initial
value of $\sigma _{infl}$ turns out to be $\sigma _{infl_{0}}\simeq 3.45,$
for $\mu =3\times 10^{-3}$ and $\beta =0.044$ the value
 $\sigma _{infl_{0}}\simeq 6$ is obtained
whereas for $\mu =10^{-3}$ and $\beta =0.027,$ $\sigma _{infl_{0}}\simeq
9.05$ \footnote{$ $In order to lower the relatively high value of 
$\sigma_{infl_{0}}$ for ``small'' $\mu$ we could increase $\chi_{0}$
to compensate the decrease of the initial F-term potential energy density
with an increase of the D-term one and choose a somewhat smaller  
$\xi_{1}$ to suppress the quantum fluctuations during the ``chaotic'' D-term
inflation. Taking, e.g, $\mu =10^{-3}$, $\beta =0.027$ and  
$\xi_{1} = \frac{1}{\sqrt{6}}$ we obtain $\sigma_{infl_{0}} \simeq 3$ if
$\chi_{0}=1.42.$ }.  
Thus, our scenario allows for a quite natural starting point
involving field values which are neither very small nor very large and an
initial energy density of order unity equally partitioned into kinetic and
potential.

We conclude by summarizing our results. In the presence of fields which
without appearing in the superpotential acquire large vevs the inflaton mass
in the context of supergravity hybrid inflation receives additional
contributions. These contributions allow the construction of new and simple
hybrid inflationary scenarios in supergravity with non-compact K$\ddot{a}$%
hler manifolds like products of coset spaces $SU(1,1)/U(1)$ which often
occur in superstring compactifications. A particularly simple model with
quite natural initial conditions allows for a detectable gravitational wave
contribution to the microwave background anisotropy.

\acknowledgments

This research was supported in part by EU under TMR contract
ERBFMRX-CT96-0090. The author would like to thank A. Kehagias for useful
discussions.

\newpage

\bigskip $
\begin{array}{ccccccc}
\mu & \beta & \sigma _{H} & \sigma _{infl_{H}} & n & n_{COBE} & r \\ 
&  &  &  &  &  &  \\ 
7.3\times 10^{-3} & 0.033 & 1.1411 & 1.5810 & 1.177 & 1.237 & 7.2\times
10^{-2} \\ 
7.0\times 10^{-3} & 0.034 & 1.1082 & 1.4914 & 1.168 & 1.222 & 6.0\times
10^{-2} \\ 
6.8\times 10^{-3} & 0.035 & 1.0817 & 1.4253 & 1.163 & 1.212 & 5.3\times
10^{-2} \\ 
6.5\times 10^{-3} & 0.036 & 1.0421 & 1.3345 & 1.155 & 1.197 & 4.3\times
10^{-2} \\ 
6.0\times 10^{-3} & 0.038 & 0.9643 & 1.1776 & 1.142 & 1.175 & 3.1\times
10^{-2} \\ 
5.0\times 10^{-3} & 0.041 & 0.7810 & 0.8791 & 1.120 & 1.138 & 1.4\times
10^{-2} \\ 
4.0\times 10^{-3} & 0.043 & 0.5657 & 0.5991 & 1.104 & 1.112 & 5.8\times
10^{-3} \\ 
3.0\times 10^{-3} & 0.044 & 0.3464 & 0.3536 & 1.093 & 1.096 & 1.8\times
10^{-3} \\ 
2.0\times 10^{-3} & 0.040 & 0.1755 & 0.1764 & 1.081 & 1.081 & 3.6\times
10^{-4} \\ 
1.0\times 10^{-3} & 0.027 & 0.06633 & 0.06638 & 1.054 & 1.054 & 2.3\times
10^{-5}
\end{array}
$

\bigskip

\bigskip

Table 1. The values of $\beta $, $\sigma _{H}$, $\sigma _{infl_{H}}$, $n$, $%
n_{COBE}$ and $r$ as functions of $\mu $ for $N_{d}=1.$

\bigskip

\bigskip

\bigskip

\bigskip \medskip

$
\begin{array}{ccccccc}
N_{d} & \beta & \sigma _{H} & \sigma _{infl_{H}} & n & n_{COBE} & r \\ 
&  &  &  &  &  &  \\ 
0 & 0.079 & 0.8528 & 0.9869 & 1.221 & 1.271 & 7.1\times 10^{-2} \\ 
1 & 0.033 & 1.1411 & 1.5810 & 1.177 & 1.237 & 7.2\times 10^{-2} \\ 
2 & 0.026 & 1.1937 & 1.7469 & 1.171 & 1.233 & 7.2\times 10^{-2} \\ 
9 & 0.0125 & 1.3031 & 2.2603 & 1.160 & 1.225 & 7.3\times 10^{-2} \\ 
16 & 0.0086 & 1.3366 & 2.5232 & 1.157 & 1.222 & 7.3\times 10^{-2}
\end{array}
$

\medskip

\bigskip

Table 2. The values of $\beta $, $\sigma _{H}$, $\sigma _{infl_{H}}$, $n$, $%
n_{COBE}$ and $r$ as functions of $N_{d}$ for $\mu =7.3\times 10^{-3}.$\qquad

\qquad

\end{document}